\documentclass[aps,pra,twocolumn,superscriptaddress,shownopacs]{revtex4}

\usepackage{graphicx}
\usepackage{hyperref}
\usepackage{amsmath}
\usepackage{epstopdf}
\usepackage{bbm}
\newcommand{\ket}[1]{\vert #1 \rangle}
\usepackage{xcolor}
\usepackage{ulem}
\usepackage{upgreek}

\begin{document}

\title{Highly-Efficient Quantum Memory for Polarization Qubits \\in a Spatially-Multiplexed Cold Atomic Ensemble}

\author{Pierre Vernaz-Gris\footnotemark[4]\footnotetext{\footnotemark[4]These authors contributed equally to the work.}}
\affiliation{Laboratoire Kastler Brossel, Sorbonne Universit\'e, CNRS, PSL Research University, Coll\`ege de France, 4 place
Jussieu, 75005 Paris, France}
\affiliation{Centre for Quantum Computation and Communication Technology, Research School of Physics and Engineering, The Australian National University, Canberra, ACT 2601, Australia}
\author{Kun Huang\footnotemark[4] }
\email{khuang@usst.edu.cn}
\affiliation{Laboratoire Kastler Brossel, Sorbonne Universit\'e, CNRS, PSL Research University, Coll\`ege de France, 4 place
Jussieu, 75005 Paris, France}
\author{Mingtao Cao}
\affiliation{Laboratoire Kastler Brossel, Sorbonne Universit\'e, CNRS, PSL Research University, Coll\`ege de France, 4 place
Jussieu, 75005 Paris, France}
\author{Alexandra S. Sheremet}
\affiliation{Laboratoire Kastler Brossel, Sorbonne Universit\'e, CNRS, PSL Research University, Coll\`ege de France, 4 place
Jussieu, 75005 Paris, France}
\author{Julien Laurat}
\email{julien.laurat@upmc.fr}
\affiliation{Laboratoire Kastler Brossel, Sorbonne Universit\'e, CNRS, PSL Research University, Coll\`ege de France, 4 place
Jussieu, 75005 Paris, France}

\date{\today}

\begin{abstract}
Quantum memory for flying optical qubits is a key enabler for a wide range of applications in quantum information science and technology. A critical figure of merit is the overall storage-and-retrieval efficiency. So far, despite the recent achievements of efficient memories for light pulses, the storage of qubits has suffered from limited efficiency. Here we report on a quantum memory for polarization qubits that combines an average conditional fidelity above 99\% and an efficiency equal to $(68\pm 2)\%$, thereby demonstrating a reversible qubit mapping where more information is retrieved than lost. The qubits are encoded with weak coherent states at the single-photon level and the memory is based on electromagnetically-induced transparency in an elongated laser-cooled ensemble of cesium atoms, spatially multiplexed for dual-rail storage. This implementation preserves high optical depth on both rails, without compromise between multiplexing and storage efficiency. Our work provides an efficient node for future tests of quantum network functionalities and advanced photonic circuits.
\end{abstract}

\maketitle

Quantum memories enabling the storage of an input photonic qubit and its later retrieval with a fidelity beating any classical device constitute essential components in quantum communication networks and optical quantum information processing \cite{Lvovsky2009,Heshami2016}. Over the past years, storage of optical qubits has been demonstrated in a variety of physical platforms, including individual atoms in high-finesse cavities \cite{Specht2011}, ion-doped crystals \cite{Gundogan2012,Clausen2012, Zhou2012} and large ensembles of neutral atoms \cite{Lukin2003,Hammerer2010,Sangouard2011}. Quantum memories capable of storing qubits encoded into multiple degrees of freedom of light have also been achieved recently \cite{Parigi}. 

The storage-and-retrieval efficiency of such devices is a stringent parameter for the envisioned applications \cite{Kimble,Walmsley,Bussieres2013,Heshami2016} and boosting this parameter has been a long-standing quest. This figure of merit is crucially important to reduce the entanglement distribution time in quantum repeater architectures and thereby develop scalable communication links \cite{Briegel,Sangouard2011,Jiang2016}. It is also essential for increasing the success rate of gate operations \cite{Chen2013} or for building up iterative quantum state engineering schemes in optical quantum circuits \cite{Felinto}. A memory efficiency exceeding the important 50\% threshold would as well enable protocols to perform in the no-cloning regime without post-selection \cite{Grosshans} or error correction for qubit losses in linear optics quantum computation \cite{Varnava}. However, to date, the highest storage-and-retrieval efficiencies achieved for qubits, independently of the photonic degrees of freedom, are below 30\% \cite{Clausen2011,Lettner2011,Kalb2015,Ding2015,Parigi,Zhou2015}. 

These limited values contrast with the recent progresses achieved in the demonstrations of optical memories. Ultra-high optical depths (OD) have indeed been obtained in laser-cooled elongated ensembles of neutral atoms \cite{Sparkes2013, Hsiao2014} and then used to realize high-efficiency single-mode optical storage \cite{Cho2016, Hsiao2016} based on long-lived collective excitations. However the increase in optical depths, which is a strong prerequisite for large efficiency \cite{Gorshkov2007}, often comes at the expense of spatial multimode capacity as the atomic cloud is elongated along a direction and radially compressed. It results in a reduced transverse size that may render some strategies such as dual-rail storage arduous. Qubit storage with large efficiency remains thereby a challenging goal. 

Here, we demonstrate a faithful quantum memory for polarization qubits with a storage-and-retrieval efficiency close to 70\%. Our realization is based on electromagnetically-induced transparency (EIT) in a single spatially-multiplexed ensemble of cold cesium atoms featuring a large optical depth. The qubits are implemented using attenuated coherent states at the single-photon level.  The reported efficiency approaches the maximal performance achievable on the D$_2$ line used here, as shown by a comprehensive model that includes all the involved atomic transitions. Relative to previous works, this advance has been made possible by combining a high-OD medium, efficient spatial multiplexing and low-noise operation.

\begin{figure*}[t!]
\includegraphics[width=1.94\columnwidth]{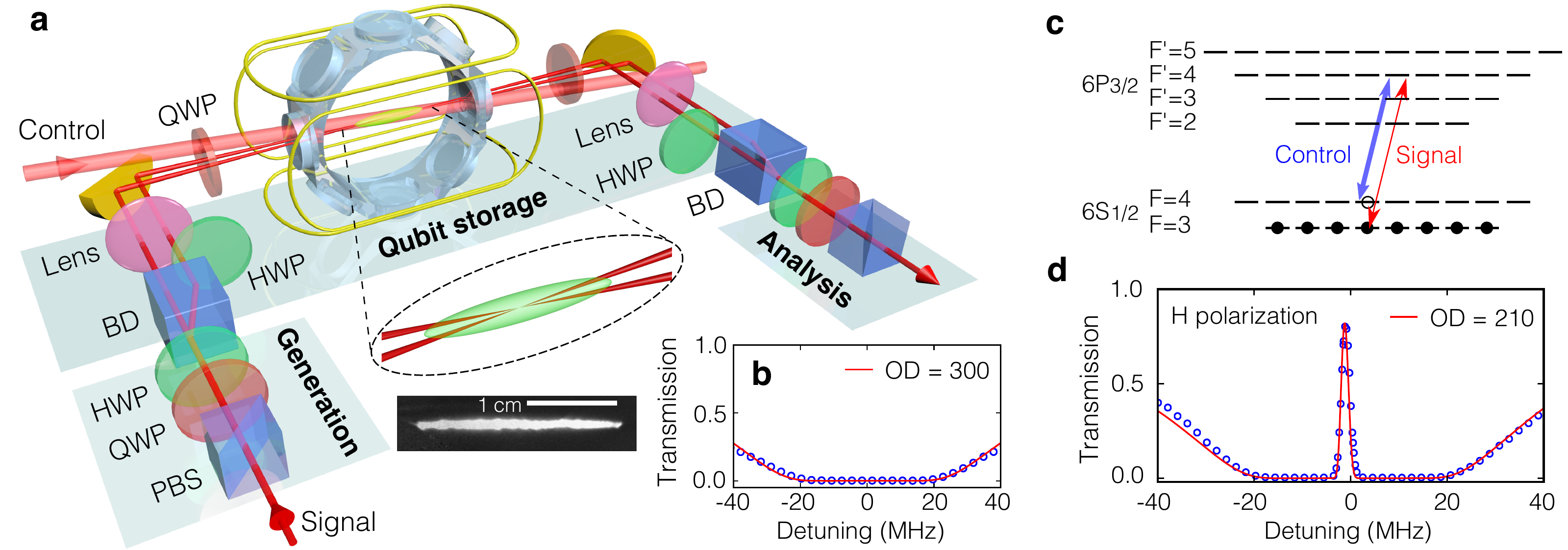}
\caption{\textbf{Quantum memory for polarization qubits in a multiplexed large-OD atomic cloud.} (a) A polarization qubit is encoded via a quarter (QWP) and a half wave plate (HWP) and converted into a dual-rail qubit with a beam displacer (BD). The orthogonally polarized beams,  separated by 4 mm, are then mapped into an elongated ensemble of laser-cooled cesium atoms prepared in a 2D magneto-optical trap in a glass chamber. The spatial multiplexing is realized by focusing the two parallel paths into the 2.5-centimeter-long ensemble with a small crossing angle of 0.5$^{\circ}$ in order to preserve a large OD for each mode, an essential but challenging feature. A single control beam propagates with an angle of 1$^{\circ}$ relative to the signal modes in the plane of symmetry. (b) A large optical depth of 300 is obtained. The blue points correspond to the experimental data while the red solid line gives the theoretical fit. (c) Energy levels of the Cs D$_2$ line involved in the EIT scheme. The atoms are prepared in $F=3$ and populate all the Zeeman levels. Signal and control fields have the same circular polarization to avoid residual absorption. A comprehensive model is derived to take into account all the atomic levels, including the excited levels out of resonance. This model allows to understand the fundamental limits for storage and retrieval in such a setting, as described in the text. (d) Typical EIT spectrum as a function of the signal detuning when the control beam is kept on resonance. The red solid line corresponds to the full model.}
\label{fig1}
\end{figure*}

To obtain an ensemble with large optical depth, our experiment is based on an elongated 2D magneto-optical trap (MOT) of cesium atoms \cite{Sparkes2013, Hsiao2014, Zhang2012}. As sketched in Fig. \ref{fig1}(a), the MOT relies on two pairs of rectangular-shaped coils and three retro-reflected trapping beams with a two-inch diameter and a total power of 350~mW. The resulting cigar-shaped ensemble has a length of 2.5~cm. 

The experiment cycle is completed with a repetition rate of 20~Hz (see Appendix A). After a 37.5-millisecond-long MOT loading, the optical depth is further increased by linearly ramping up the magnetic field gradient in the transverse directions to radially compress the ensemble. For this purpose, the trapping coil current is increased from 4~A to 16~A over 8~ms. After switching off the MOT coils, polarization gradient cooling is performed for 1.95~ms by ramping down the power of the trapping and repump beams following an exponential profile, while ramping up the detuning of the trapping beam from -17~MHz to -107~MHz.  

The atoms are prepared in the $\ket{g} = \ket{6S_{1/2}, F=3}$ state by turning off the repump light earlier than the trapping light, as well as sending an additional depumping light resonant on the $\ket{s} = \ket{6S_{1/2}, F=4}$ to $\ket{e} = \ket{6P_{3/2}, F'=4}$ transition. This transfer is required as no perfect EIT can be obtained in the absence of Zeeman optical pumping if $\ket{s}$  is used as the initial ground state. Finally, experiments are started 2 ms after the MOT coils are turned off to allow sufficient decay of the magnetic field. Three pairs of coils are used to compensate residual magnetic fields and to limit the inhomogeneous broadening to 50 kHz, as measured by microwave spectroscopy.

At the end of the preparation stage, the OD for a probe resonant to the $\ket{g} \to \ket{e}$ transition reaches a value of about 300, as shown in Fig. \ref{fig1}(b). The temperature of the atoms is measured to be 20 $\upmu$K with a time-of-flight technique.

\begin{figure}[t!]
\includegraphics[width=0.94\columnwidth]{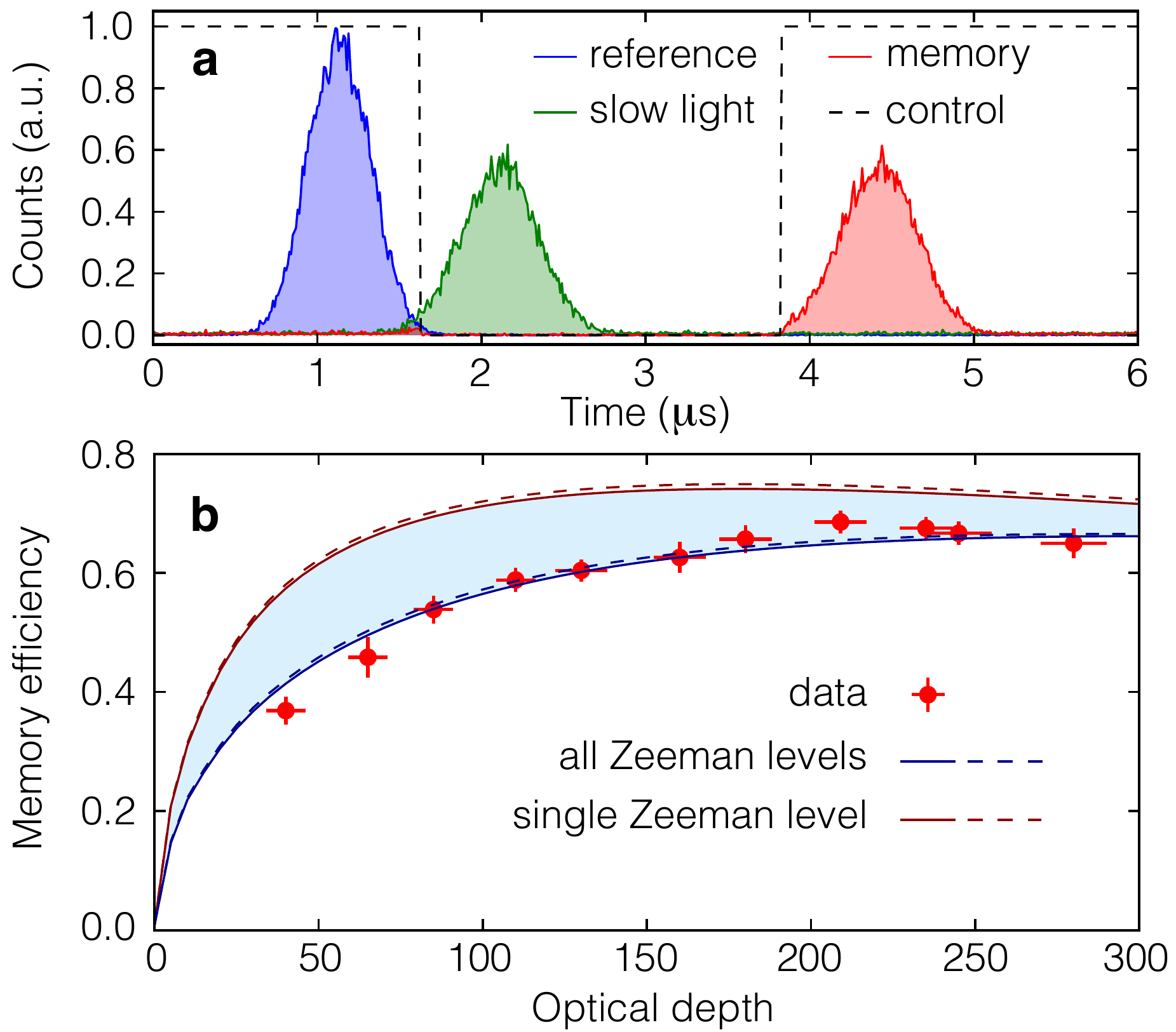}
\caption{\textbf{High-efficiency storage and retrieval.} (a) Histogram of the photodetection counts. The blue-filled region gives the reference pulse without atoms while the green and red regions correspond to the slow and the stored-and-retrieved pulses, respectively. The black dashed line indicates the control intensity for the storage experiment. The memory efficiency reaches $(69\pm1)\%$. (b) Storage-and-retrieval efficiency as a function of the optical depth of the atomic cloud. The theoretical lines correspond to the two limiting cases with a single Zeeman level and all Zeeman levels, respectively. For each case, the solid lines correspond to an intrinsic ground state decoherence estimated to $\gamma_0 = 10^{-3} \Gamma$ while the dashed lines correspond to the limiting case with $\gamma_0 = 0$. $\Gamma$ denotes the natural linewidth of the excited state. The errors are obtained from multiple independent measurements.}
\label{fig2}
\end{figure}

Having prepared a large-OD ensemble, we now turn to the memory protocol. The reversible mapping is based on EIT that enables the conversion of a signal photon into a long-lived collective excitation by dynamically changing the power of an auxiliary control field \cite{Liu2001,Chaneliere,Eisaman,Choi2008}. The signal and control beams, which are tightly phase locked (see Appendix A), have waist diameters of 250 $\upmu$m and 2 mm respectively, and they intersect at the centre of the MOT with a small angle of 1$^{\circ}$. The control beam on the $\ket{s} \to \ket{e}$ transition has the same circular polarization as the signal, as shown in Fig. \ref{fig1}(c), to avoid absorption in an atomic system involving various EIT channels due to the presence of Zeeman sublevels. Proper alignment for the light polarization is especially important for a high-OD medium, as only a small residual fraction of OD would lead to a significant absorption for the signal. Figure \ref{fig1}(d) gives a typical EIT spectrum. A transmission close to 80\% is obtained at large OD.  

To ensure negligible leakage during the storage process, the power of the control beam is chosen to provide a slow-light delay equal to twice the probe pulse duration when the control is continuously on. Figure \ref{fig2}(a) gives a single-photon level measurement of the slowed pulse and an example of a dynamic memory operation with a few microseconds storage. Before detection, the signal passes through a home-made atomic filter and a commercial lens-based cavity for spectral filtering (Quantaser FPE001A), with an overall rejection of 70 dB for the control field.

Thanks to the large OD achieved here, we could investigate the scaling behaviour of the storage-and-retrieval efficiency. Figure \ref{fig2}(b) shows the efficiency as a function of the OD, which is varied by adjusting the power of the trapping beams during loading. At each value, the time delay is maintained constant by adapting the control beam power. As can be seen, the memory efficiency saturates at an OD of around 200 before decreasing. The maximal efficiency achieved here reaches (69$\pm$1)\%. This represents a record on the cesium D$_2$ line and, more importantly, the highest achievable value in this configuration.

To understand this scaling and OD tradeoff, the complex level structure has to be taken into account. In the alkali-metal atoms, hyperfine interaction in the excited state indeed introduces several levels and EIT features can differ from the usual three-level $\Lambda$ approximation, as previously studied in our group in various contexts \cite{Sheremet2010, Mishina2011,Scherman2011,Giner2013}. Even for cold atoms, the off-resonance excitation of multiple excited levels can have a strong effect on the medium susceptibility. This is especially true for the D$_2$ line of cesium atoms, for which the levels are only separated by 30 to 50$\Gamma$, where $\Gamma$ denotes the natural linewidth of the excited state. These off-resonant excitations result in AC Stark shifts and effective additional ground state decoherence proportional to the control power. This decoherence rate limits the achievable transparency and therefore the storage efficiency at large ODs \cite{Hsiao2016}. In addition, the atoms can be distributed in the ground state over many Zeeman sublevels, as it is the case in our experiment, and this configuration can lead to further inefficiency.

On Fig. \ref{fig2}(b), experimental results are compared with a full model based on the Maxwell-Bloch equations and that takes into account the interaction of the probe and control field not only with all the excited levels but also with the Zeeman states (see Appendix C). The theoretical lines correspond to the case with a single Zeeman level of the ground state $\ket{g}$, e.g. $m=+3$, and to the case with an equal population in all the Zeeman levels. As can be seen, the experimental results are in strong agreement with this second case. The efficiency would only be slightly increased to 75\% by optical pumping, which is a  challenging task for atomic ensembles with very large OD. The model also confirms that the intrinsic ground state decoherence $\gamma_0 = (1.0 \pm 0.5)\times10^{-3} \Gamma$ estimated in our experiment is not the limiting factor for the memory performance. 

Next we present the extension of the setup to the qubit storage. The polarization mapping is implemented by using a dual-rail strategy in a single ensemble \cite{Matsukevich2004,Chou2007}. This method has been used in various implementations \cite{Laurat2007,Parigi,England2012,Kupchak2014}, but the challenge was here to maintain a large OD and a strong noise cancellation despite this multiplexing in an elongated ensemble with a very reduced cross-section. 

For this purpose, as illustrated in Fig. \ref{fig1}(a), the signal beam passes through a beam displacer based on a birefringent calcite crystal (Thorlabs BD40) that provides a large 4-mm separation between the two orthogonally polarized beams. The two paths are then focused on the centre of the MOT \cite{Ding2016} with a 500-millimeter focal length lens. Their transverse separation at the MOT edge is estimated to be 100 $\upmu$m, which is much smaller than the 1-millimeter transverse size of the MOT. In contrast to the usual parallel scheme, i.e. without focusing the signal beam, here both beams cross the centre of the MOT, enabling the attainment of a large and similar OD for the two paths. In the experiment, the OD is chosen to be around 200. 

At the memory output, the two paths are recombined into a single spatial mode with a second beam displacer. The position of the second lens is optimized to obtain a visibility over 99\% between these two paths, which is a critical step to achieve high fidelity. The two displacers form a passively stable Mach-Zehnder interferometer, where the relative phase is set to zero by adjusting the tilt of the second beam displacer.

\begin{figure}[t!]
\includegraphics[width=0.99\columnwidth]{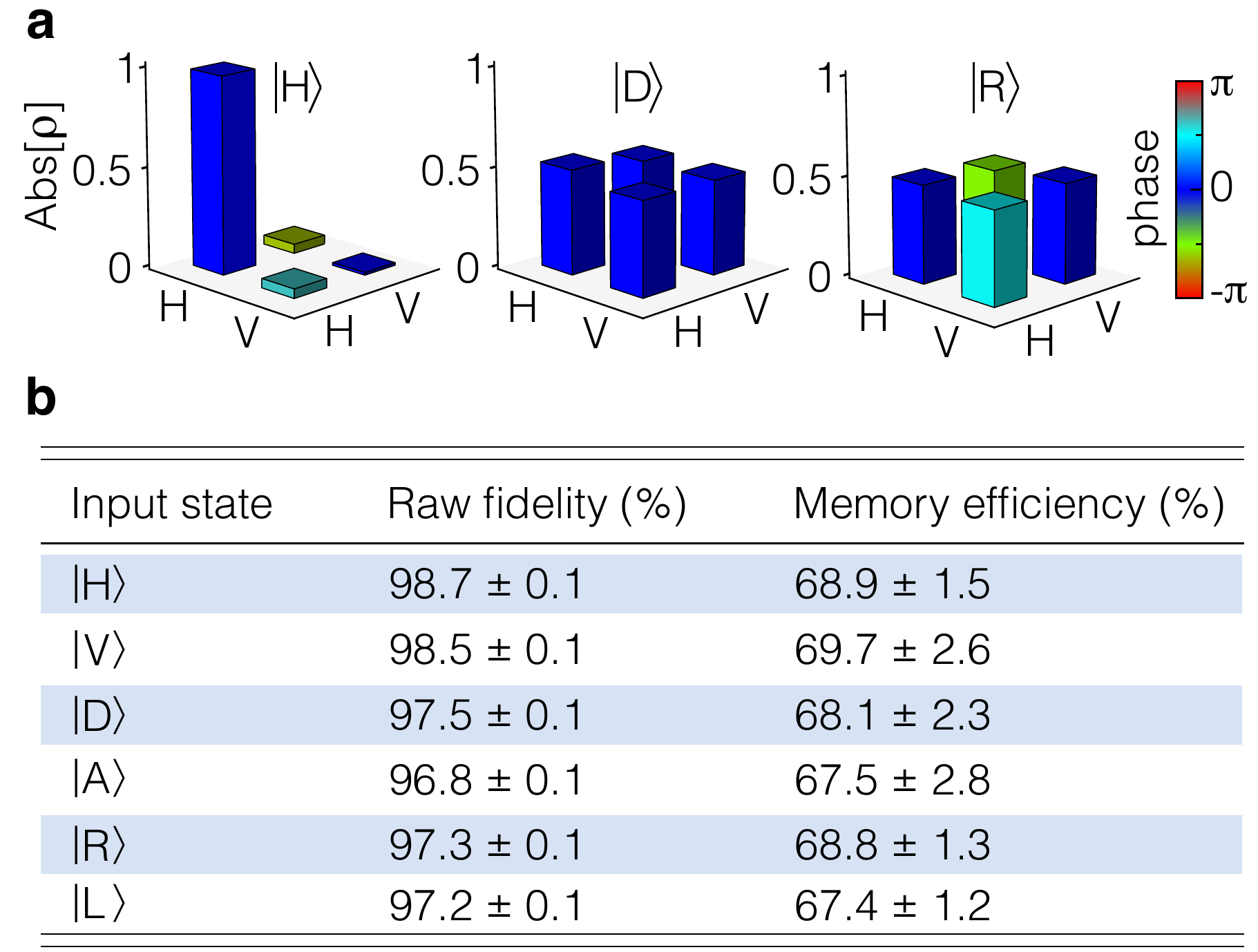}
\caption{\textbf{Quantum state tomography of the retrieved polarization qubits.} (a) Reconstructed density matrices for the retrieved states after a 1.2-$\upmu$s storage time. The height of the bar represents the absolute value while the colour denotes the phase. No background has been subtracted. (b) Conditional fidelities and memory efficiencies for the set of six input qubits. The error bars for the fidelity are estimated by taking into account the statistical uncertainty of photon counts. The error for the efficiency is obtained from multiple measurements. The mean number of photons per pulse is $\bar{n} = 0.5$.}
\label{fig3}
\end{figure}

We now proceed to the qubit storage in this dual-rail setting. Polarization qubits are implemented with weak coherent states with a mean photon number per pulse $\bar{n} = $ 0.5 and subsequently stored into the memory. The retrieved states are then characterized by usual quantum state tomography \cite{James2001}. Figure \ref{fig3}(a) give the reconstructed density matrices in the $\{\ket{H}, \ket{V}\}$ logical basis. From the measured matrices, one can estimate the conditional fidelity of the output states with the initially encoded state. The values for the complete set of inputs are listed in Fig. \ref{fig3}(b). The average fidelity is $(97.7\pm0.8)\%$ and raises up to $(99.5\pm0.5)\%$ after correction for background noise that mainly comes from residual control leakage and detector dark counts. In the absence of an input signal, the background floor corresponds to $5\times10^{-4}$ events per detection window.

\begin{figure}[b!]
\includegraphics[width=0.94\columnwidth]{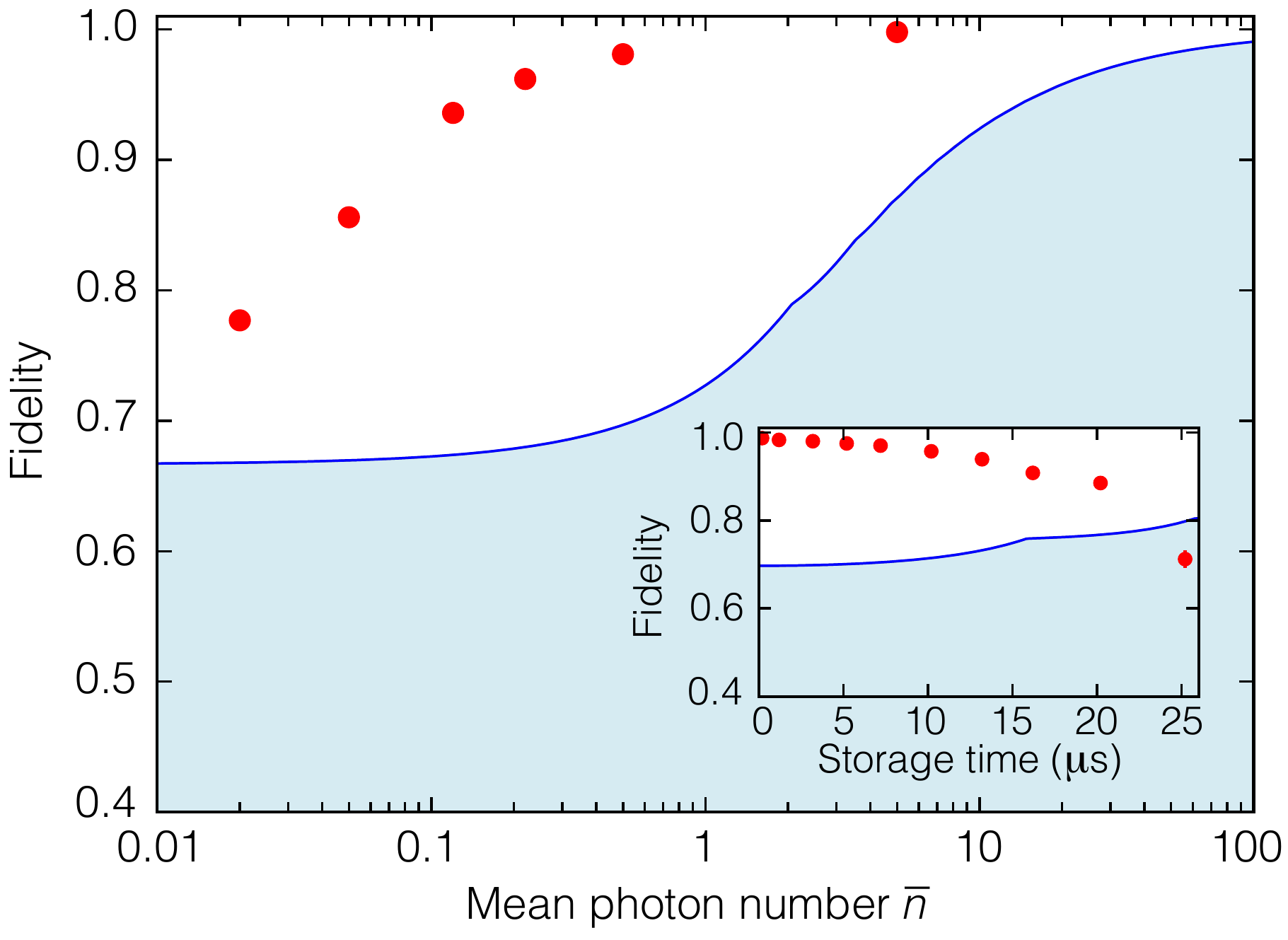}
\caption{\textbf{Storage and retrieval beyond classical benchmark}. The fidelity is given as a function of the mean photon number per pulse $\bar{n}$, for a 1-$\upmu$s storage time. The inset shows the fidelity as a function of the storage time, for a mean photon number $\bar{n}=0.5$. No background correction has been applied. The blue solid line  indicates the classical limit for the finite storage-and-retrieval efficiency and takes into account the Poissonian statistics of the weak coherent states. The error bars are smaller than the data points and are given by the standard deviations of fidelities for the set of stored states.}
\label{fig4}
\end{figure}

In order to conclude about the quantum performance of the storage, we need to compare our fidelity with the maximal one achievable using a classical memory device, based for instance on the so-called measure-and-prepare strategy. It can be shown that the classical benchmark is given by a fidelity equal to $(N+1)/(N+2)$ for a state containing $N$ photons \cite{Massar1995}, which is equal to  2/3 for the particular case of a single photon. In our case, the bound has to be modified to take into account the Poissonian statistics of the probe state and the finite memory efficiency, as done in Refs. \cite{Specht2011, Gundogan2012}. Figure \ref{fig4} presents the achieved fidelities as a function of the mean photon number per pulse $\bar{n}$. The measured fidelities are largely above the classical benchmark for $\bar{n}$ as low as 0.02. As shown in the inset of Fig. \ref{fig4} , the memory time for maintaining the quantum nature of the storage reaches more than 20 $\upmu$s. In our experiment, the lifetime is mainly limited by the residual magnetic field along the elongated atomic cloud (see Appendix B). 

For these measurements, the average storage-and-retrieval efficiency reaches $(68.5\pm2)\%$, as detailed in Fig. \ref{fig3}(b). This value is the highest efficiency reported so far for a reversible memory that demonstrates the quantum storage of photonic qubits. Moreover, we have shown that, because of the multiple level structure of the cesium D$_2$ line, the achieved efficiency is the best one can obtain in this configuration. Direct extension of our setup to the D$_1$ line where excited levels are much more separated should enable to reach an efficiency above 90\% \cite{Hsiao2016}, although this has yet to be demonstrated in the quantum regime. 

In our realization, the pulse duration of the qubits has been chosen around 400~ns to obtain the highest efficiency after one pulse-width delay. It corresponds to a memory bandwidth of a few MHz, as expected from EIT storage \cite{Heshami2016}. Single-photon sources with sub-MHz bandwidths have been demonstrated \cite{Du2008,Fekete2013,Rambach2016} and can be adapted to our reported quantum memory.

In summary, we have demonstrated a highly-efficient memory for optical qubits by successfully operating a large-OD elongated atomic ensemble in a dual-rail configuration. This combination enables the reversible mapping of arbitrary polarization states not only with fidelities well above the classical benchmark but also with an overall storage-and-retrieval efficiency close to 70\%. This value represents the highest efficiency to date for the storage and readout of optical qubits in any physical platform and is more than double of the previously reported values. It also outperforms the important 50\% threshold required to beat the no-cloning limit without post-selection. 

Besides the aforementioned network architecture scalability and potential loss-tolerant schemes, the achieved efficiency opens the way to first tests of advanced quantum networking tasks where the storage node efficiency plays a critical role, such as in certification protocols or unforgeable quantum money \cite{Nori,Bozzio}. Moreover, the designed platform is directly compatible with recent works based on spatially structured photons and multiple-degree-of-freedom storage \cite{Parigi} and can now yield very efficient realizations to boost high-capacity network channels.

\begin{acknowledgments}
We thank V. Parigi and C. Arnold for their contributions in the early stage of the experiment. This work was supported by the European Research Council (ERC Starting Grant HybridNet). M.C. and A.S.S. are supported by the EU (Marie Curie Fellowship). J.L. is a member of the Institut Universitaire de France. 
\end{acknowledgments}

\clearpage

\onecolumngrid
\appendix

\section{Laser sources and experimental timing}

Two main laser sources are used in the experiment. A Ti:sapphire laser (MSquared, SolsTiS) is stabilized on a reference cavity and frequency-locked via saturated absorption. It is used for seeding a tapered amplifier (Toptica Photonics, BoosTA) to deliver 350~mW for trapping beams. This laser source is also used for the control beam in the EIT protocol. The second laser source is an external-cavity diode (Toptica Photonics, DL Pro). It provides the 10-mW repumper and is also used for the probe. The two laser sources are phase-locked at the cesium hyperfine splitting frequency (Vescent Photonics, D2-135). Frequency tunings are realized via acousto-optical modulators in double-pass configuration. 

\begin{figure}[b!]
\def\thefigure{{S1}}
\centering
\vspace{0.5cm}
\includegraphics[width=0.85\columnwidth]{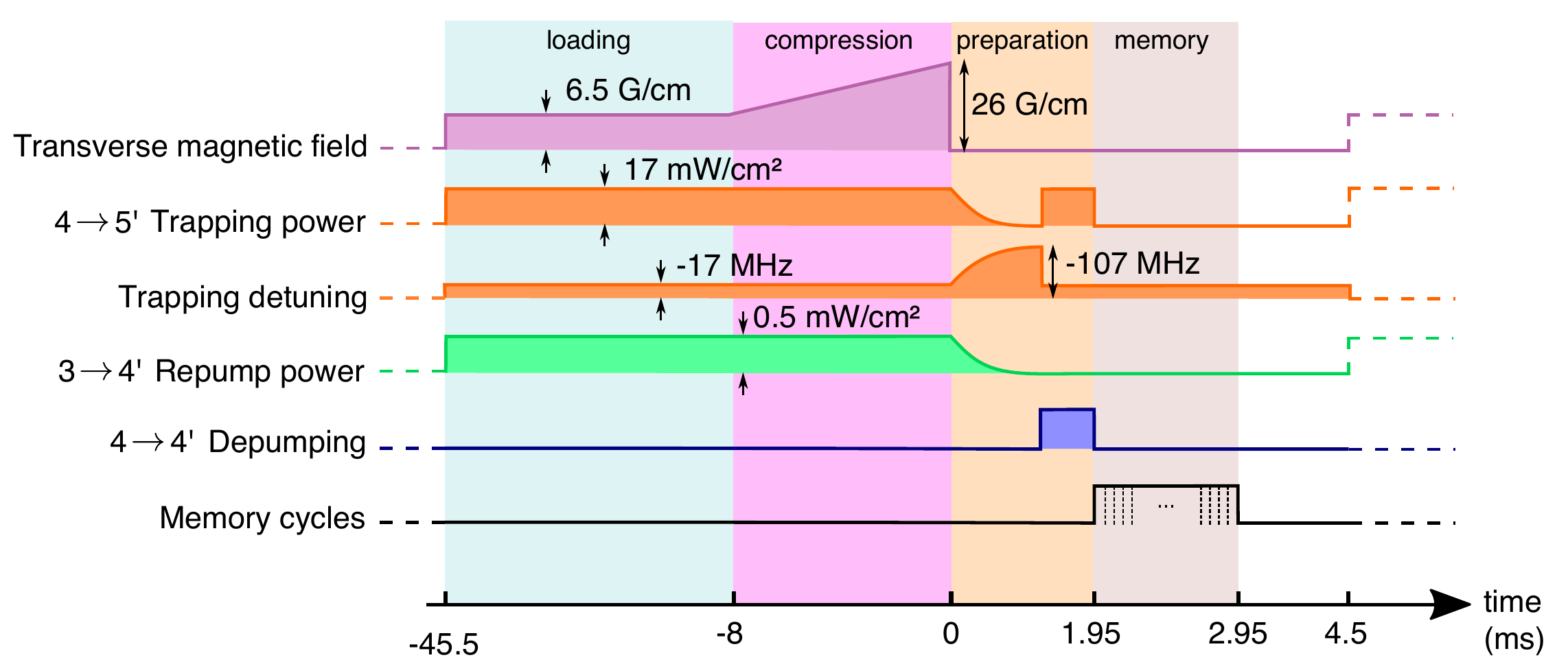}
\caption{Timing diagram of the experiment.}
\label{timing}
\end{figure}

The timing is shown in  Figure \ref{timing}. The experiment is performed at a repetition rate of 20 Hz. The two pairs of coils (dimensions 270~mm $\times$ 110~mm) are driven by two independent power supplies (Delta Elektronika, SM52-30). After a trap loading phase of 37.5 ms, the atomic cloud is compressed by linearly ramping the current supply from 4 A to 16 A. The trapping magnetic field is then turned off ($t=0$) via two home-made electronic switches. A preparation phase follows, with the trapping and repump powers ramped down while the trapping detuning is increased from -17 MHz to -107 MHz in 1 ms. Before the end of the preparation process, a 950-$\upmu$s resonant (F=4 $\rightarrow$ F'=4) pulse and a 10-MHz red-detuned trapping beam are sent together to transfer the atoms to the F=3 ground state. The trapping coils current is sinked with an exponential decay time of about 50 $\upmu$s. However, the induced eddy currents of the surrounding metallic components have a longer decay time. To reduce their perturbations, the residual magnetic field cancellation via three pairs of coils is optimized 2 ms later after the switching. Memory operations are then performed during a period of 1 ms. Depending on the storage time, this interval is split into 25 to 100 repetitions. Photons are detected by a single avalanche photodiode (SPCM-AQR-14-FC) and recorded with a FPGA-based digitizer with a time resolution of 10 ns. For each polarization projection, 10$^5$ memory sequences are accumulated.

\section{Memory lifetime and decoherence}
Three decoherence mechanisms can be evaluated independently. First, the atomic motion related to the finite temperature results in a possible loss of the atoms from the interaction area. The thermal velocity of an atom of mass $m$ at a temperature $T$ is given by $v = \sqrt {{{{k_\textrm{B}}T} \mathord{\left/
 {\vphantom {{{k_\textrm{B}}T} m}} \right.
 \kern-\nulldelimiterspace} m}}$. The transit time can therefore be estimated by ${\tau _1} = {D \mathord{\left/
 {\vphantom {D v}} \right.
 \kern-\nulldelimiterspace} v}$ with $D\simeq250~\upmu$m the diameter of the probe beam. With a  temperature $T$ estimated at 20 $\upmu$K by a time-of-flight measurement, the corresponding transit time is ${\tau _1} = 7$~ms. This is not a limiting factor in our experiment.
 
The two other decoherence contributions come from the dephasing of the collective excitation. The first source of possible dephasing is the so-called motional dephasing due to the strong angular dependence of EIT \cite{Carvalho2004,Zhao09}. In the experiment, we used indeed an off-axis configuration and the control and probe beams are overlapped with an angle $\theta\simeq1^\circ$. The resulting lifetime is then given by:
\begin{equation}
\tau_2  = \frac{\lambda }{{2\pi \sin \theta }}\sqrt {\frac{m}{{{k_\textrm{B}}T}}}.
\label{eq2}
\end{equation}
In our case, this expression leads to a decay time ${\tau _2} = 220~\upmu$s. The second dephasing process is caused by residual magnetic fields which result in an atom-dependent Larmor precession. The magnetic field is compensated via three pairs of coils and the inhomogeneous broadening in the ground state is therefore limited to around 50~kHz, as measured via microwave spectroscopy. By assuming a gradient of magnetic field equal to 8 mG/cm along the length of the ensemble $L=2.5$~cm and a Gaussian atomic distribution, the expected time constant is ${\tau _3}=15~\upmu$s \cite{Choi2011}. This dephasing is the main decoherence source in our experiment.

\begin{figure}[htpb!]
\def\thefigure{{S2}}
\begin{minipage}[b]{0.35\columnwidth}
\centering
\caption{Retrieval efficiency as a function of the storage time. The blue points give the experimental measurements while the red solid line corresponds to the full model explained in Appendix C, with a measured OD equal to 200 and a 50~kHz inhomogeneous broadening. The  shaded area corresponds to an uncertainty on the residual magnetic field about $\pm1$ mG.cm$^{-1}$. Errors were estimated assuming Poissonian statistics.}
\vspace{2.7cm}
\label{efficiency_time}
\end{minipage}
\hspace{0.15cm}
\begin{minipage}[b]{0.55\linewidth}
\centering
\includegraphics[width=\textwidth]{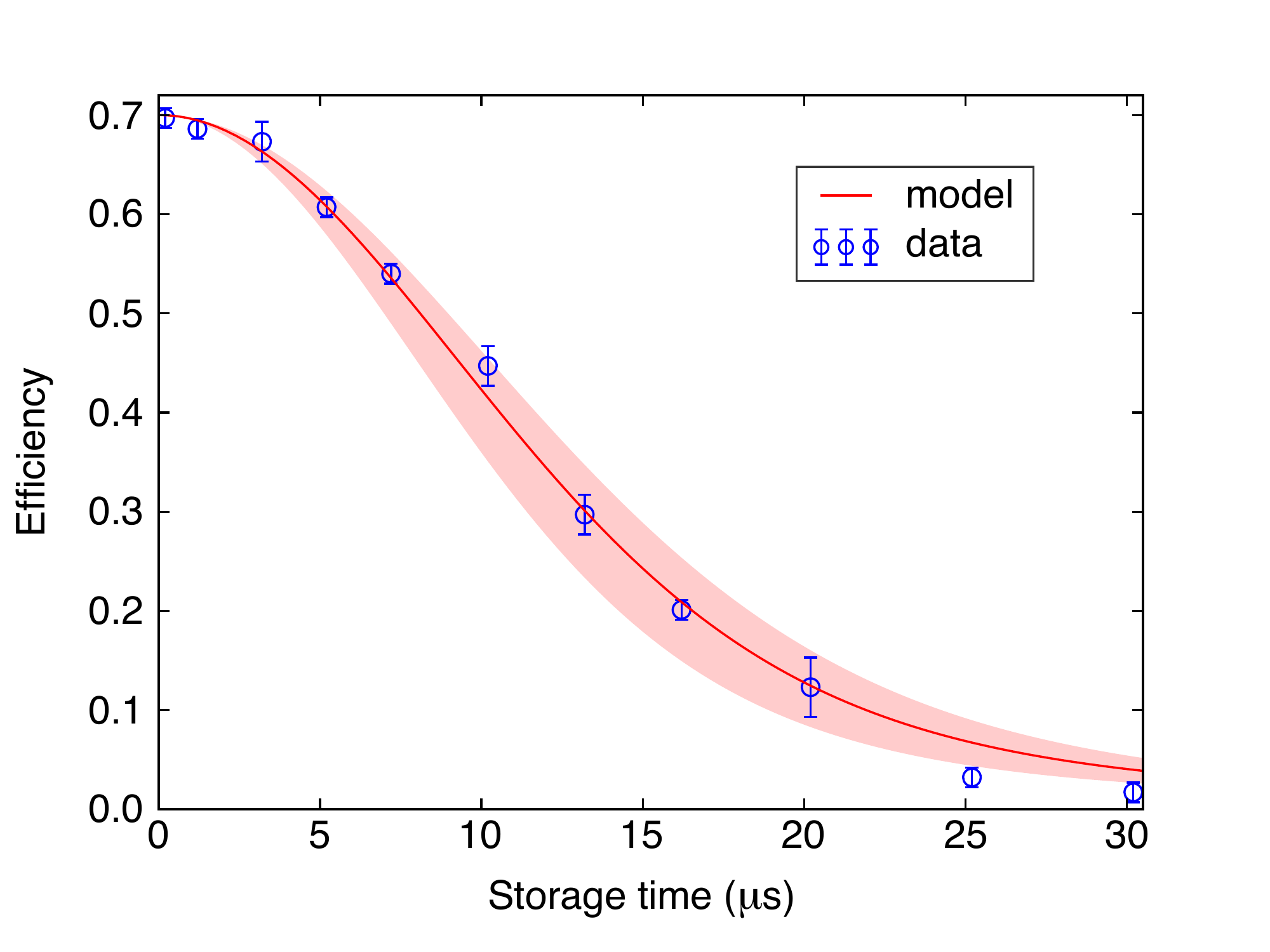}
\end{minipage}
\end{figure}

Figure \ref{efficiency_time} gives the retrieval efficiency as a function of the storage duration. The solid line corresponds to the full model derived in the following section.

\section{Multi-level theoretical model for light storage and retrieval}

In our theoretical description, we consider the interaction of the probe and control fields with the full D$_2$-line structure of $^{133}$Cs atoms. The influence of the multi-level structure in alkali-metal atoms, beyond the three-level $\Lambda$ approximation or double-$\Lambda$ model, was studied theoretically and experimentally in our group in various contexts, e.g. electromagnetically-induced transparency (EIT) with broadened transitions \cite{Mishina2011,Scherman2012}, Raman configuration \cite{Sheremet2010} or EIT/ATS transition with cold atoms \cite{Giner2013}. Even for cold atoms, the off-resonance excitation of multiple excited levels can have a strong effect on the medium susceptibility and therefore on the memory efficiency. In addition to all the excited levels, we also include here the Zeeman sublevels to compare implementations with and without optical pumping. 

\subsection{Level scheme: multiple excited levels and Zeeman sublevels}
\begin{figure}[t!]
\def\thefigure{{S3}}
\begin{minipage}[b]{0.35\columnwidth}
\centering
\caption{Level scheme for $^{133}$Cs D$_2$-line. The $\sigma^{+}$-polarized control field couples the $|F_s = 4, m\rangle \rightarrow |F'_e = 4, m + 1\rangle$ transitions, while the $\sigma^{+}$-polarized probe field is scanned near resonance with $|F_g = 3, m\rangle \rightarrow |F'_e = 4, m + 1 \rangle$ transitions.}
\vspace{1.7cm}
\label{fig_SM1}
\end{minipage}
\hspace{0.25cm}
\begin{minipage}[b]{0.55\linewidth}
\centering
\includegraphics[width=\textwidth]{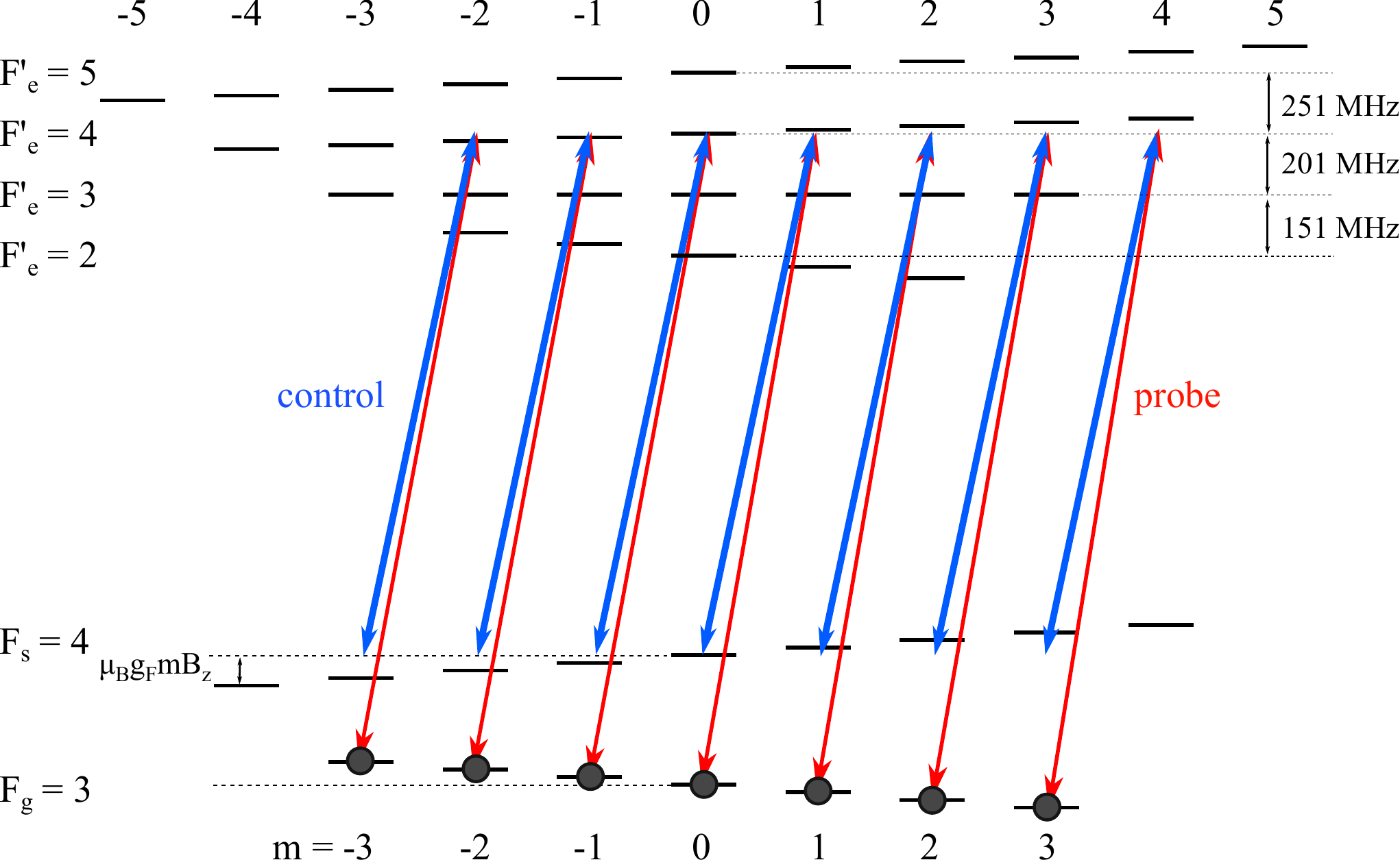}
\end{minipage}
\end{figure}

The level scheme is shown in Figure \ref{fig_SM1}. Atoms equally populate the Zeeman sublevels $|g_m\rangle = |6S_{1/2}, F = 3, m\rangle$, which are coupled to the excited states $|e_{F',n}\rangle = |6P_{3/2}, F' = 2,3,4,  n = m + 1 \rangle$ by a weak $\sigma^{+}$-polarized probe field. The $\sigma^{+}$-polarized control field couples the second ground state $|s_{m}\rangle = |6S_{1/2}, F = 4, m\rangle$ with the excited levels $|e_{F',n}\rangle = |6P_{3/2}, F' = 3,4,5, n = m + 1\rangle$. The atomic medium is optically thick for the probe field. Such polarization scheme with atoms initially in $F_g=3$ provides interaction of the probe with the atomic ensemble without residual absorption in the absence of optical pumping. The detunings of the probe field with frequency $\omega_p$ and the control field with frequency $\omega_c$ from the atomic transitions $|g_{m}\rangle \rightarrow |e_{F' = 4, n}\rangle$ and $|s_{m}\rangle \rightarrow |e_{F' = 4, n}\rangle$ in the absence of a magnetic field are noted as $\Delta_p = \omega_p - \omega_{e_{F',n}g_{m}}$ and $\Delta_c = \omega_c - \omega_{e_{F',n}s_{m}}$ respectively.

\subsection{Residual magnetic field}
In this model, we also include a residual magnetic field, which is the main decoherence source in our implementation. We consider a magnetic field gradient along the $z$ axis with $B(z) = B_0 z$. This magnetic field $B(z)$ results in the splitting of Zeeman sublevels in the ground and excited states and leads to additional detuning for the probe and control fields from magnetically-insensitive transitions:
 \begin{equation}
 \delta_{m}^{(p,c)}(z) = \frac{\upmu_\textrm{B} \left[mg_{F_{g,s}} - (m + 1)g_{F'_e}\right] B(z)}{\hbar},
 \end{equation}
 where $\upmu_\textrm{B}$ is the Bohr magneton and $g_{F_g}, g_{F_s}, g_{F'_e}$ are the hyperfine Land\'e factors for $F_g$, $F_s$ and $F'_e$ respectively.  
 
\subsection{Optical Bloch equations}

The atomic system evolution is described by:
\begin{equation}
\frac{d\hat{\rho}}{dt} = \frac{i}{\hbar}\left[\hat{\rho}, \hat{H} \right] -\hat{\Gamma}\hat{\rho}.
\label{Bloch}
\end{equation}
Here $\hat{\Gamma}$ is a relaxation operator describing the radiative decay rate of the excited states and the decoherence processes in the ground state.

The Hamiltonian of the system can be written in the following form:
\begin{equation}
\hat{H} = \hat{H}_0 + \hat{V} = \hat{H}_{\textrm{atom}} + \hat{H}_{\textrm{field}} + \hat{V}_E+\hat{V}_B,
\label{Hamiltonian}
\end{equation}
where the non-perturbative part of the Hamiltonian $\hat{H}_0$ is given by the sum of the free Hamiltonian operators of the atoms $\hat{H}_{\textrm{atom}}$ and the electromagnetic field $\hat{H}_{\textrm{field}}$. The dipole interaction part is written in the rotating-wave approximation and consists of the dipole interactions between the atom and the probe and control fields:
\begin{equation}
\hat{V}_E = \hat{V}_p + \hat{V}_c,
\end{equation}
where 
\begin{equation}
\hat{V}_p = -\sum_{m = -3}^{3} \sum_{F'=2,3,4}d_{e_{F',m+1}g_m}|e_{F',m+1}\rangle \langle g_m| E_p^{(+)} +  H.c.
\end{equation}
and 
\begin{equation}
\hat{V}_c = - \sum_{m = -3}^{3}\sum_{F'=3,4,5}d_{e_{F',m+1}s_m}|e_{F',m+1}\rangle \langle s_m| E_c^{(+)} +  H.c.
\end{equation}
In the interaction picture, the positive frequency components of the electromagnetic field for the probe and the control modes  are $E_p^{(+)} = \epsilon_p \cdot e^{-i\omega_p t}$ and $E_c^{(+)} = \epsilon_c \cdot e^{-i\omega_ct}$ respectively, $d_{ij} = \langle i| \hat{\mathbf{d}}|j \rangle$ is the electric dipole moment of the atom between levels $i$ and $j$. 

The optical Bloch equations can be written for slowly-varying amplitudes of the optical coherences $\sigma_{e_{F',m+1} g_m} ~= ~\rho_{e_{F',m+1} g_m} e^{i\omega_p t}$ and $\sigma_{s_{m} g_m} = \rho_{s_m g_m} e^{i(\omega_p - \omega_c) t}$ and solved following the approach developed in \cite{Mishina2011} as: 
\begin{eqnarray}
\frac{d\sigma_{e_{F',m+1} g_m}(z)}{dt} &=& i\left(\Delta_p + \delta_{m}^{(p)}(z) - \omega_{e_{F',m+1}e_{F'=4,m+1}}(z) + i\Gamma/2 \right)\sigma_{e_{F',m+1}g_m}(z)  \nonumber\\
&&\qquad\qquad\qquad\qquad\qquad\qquad+ i\left( \frac{\Omega^{(p)}_{e_{F',m+1}g_m}}{2}\rho_{g_mg_m} + \frac{\Omega^{(c)}_{e_{F',m+1}s_m}}{2}\sigma_{s_mg_m}(z) \right)
\nonumber\\
\frac{d\sigma_{s_mg_m}(z)}{dt} &=& i\left((\Delta_p + \delta_{m}^{(p)}(z)) - (\Delta_c + \delta_{m}^{(c)}(z)) + i\gamma_0 \right)\sigma_{s_mg_m}(z) 
+ i\sum_{F'=3,4,5}\frac{\Omega^{(c)}_{s_me_{F',m+1}}}{2}\sigma_{e_{F',m+1}g_m}(z)
\label{systemEq}
\end{eqnarray}
where $\Gamma $ is the excited state decay rate, $\gamma_0$ is the ground state decoherence, $\hbar\cdot\omega_{e_{F',m+1}e_{F'=4,m+1}}(z)$ is the energy difference between the hyperfine levels $|F'_e, m + 1\rangle$ and $|F'_e = 4, m + 1\rangle$ of the excited state at position $z$, $\Omega^{(p)} = 2\mathbf{d}\cdot\mathbf{E}_p/\hbar$ and  $\Omega^{(c)} = 2\mathbf{d}\cdot\mathbf{E}_c/\hbar$ are the probe and control field Rabi frequencies respectively. The system of equations (\ref{systemEq}) is solved by taking the Fourier transform of the atomic coherences $\tilde{\sigma}_{ij}(\omega,z) = \int_{-\infty}^{\infty}\sigma_{ij}(t,z)e^{i\omega t}dt$.

The linear response of the atomic system to a weak probe field can then be described by the susceptibility given by the following expression:
\begin{equation}
\chi(\omega, z, \Delta_p, \Delta_c) = -\sum_{m = -3}^3\sum_{F' = 2,3,4}n_0(z) d_{g_m e_{F',m+1}}\tilde{\sigma}_{e_{F',m+1} g_m}(\omega, z),
\label{susceptibility}
\end{equation}
where $\tilde{\sigma}_{e_{F',m+1} g_m}(\omega,z)$ are the solutions for the slowly-varying amplitudes of the optical coherences between the ground state $|g_m\rangle$ and the excited state $|e_{F', m+1}\rangle$ addressed by the probe field. We assume an atomic cloud with a Gaussian distribution of the atomic density $n_0(z) = n_0e^{-4z^2/L^2}$, where $L$ is the length of the atomic medium.

\subsection{Stark shift and additional ground state decoherence due to the excited levels}

The non-resonant coupling of the control field with the excited states $|F'_e = 3,5\rangle$ results in a significant modification of the atomic response compared to the standard $\Lambda$-scheme approximation \cite{Mishina2011}. These couplings result in Stark shifts and an additional effective ground state decoherence:
\begin{eqnarray}
\Delta_p^{(\textrm{eff})} &=&\Delta_p + \sum_{m = -3}^{3}\sum_{F' = 3,5}\frac{|\Omega^{(c)}_{e_{F',m+1}s_m}|^2/4}{\omega_{e_{F',m+1}e_{F'=4,m+1}}},
\nonumber\\
\gamma_0^{(\textrm{eff})} &=& \gamma_0 + \sum_{m = -3}^{3}\sum_{F' = 3,5}\frac{|\Omega^{(c)}_{e_{F',m+1}s_m}|^2/4}{\omega^2_{e_{F',m+1}e_{F'=4,m+1}}}\frac{\Gamma}{2}.
\label{effective}
\end{eqnarray}
Importantly, these additional terms depend on the power of the control beam. Recently these contributions were also described in \cite{Chen2016} for N-type four-level model in the context of optical memories with atoms optically pumped to one Zeeman sublevel. In contrast, the present model includes not only the several excited levels but also the Zeeman sublevels, as they are involved in our experimental realization.

\subsection{Loss during the pulse propagation in the medium}
The loss during the pulse propagation in the medium gives the upper bound for the memory efficiency. This propagation can be described by the standard macroscopic Maxwell equation:
\begin{equation}
\left[\frac{1}{c}\frac{\partial}{\partial t} + \frac{\partial}{\partial z} \right]\epsilon(z,t) = 
2\pi i\frac{\omega}{c}\int_{-\infty}^{t}dt' \chi(z,t,t')\epsilon(z,t').
\label{Maxwell}
\end{equation}
The solution are found via the Fourier representation:
\begin{equation}
\epsilon_{\textrm{out}}(L, t) = 
\int_{-\infty}^{\infty}\frac{d\omega}{2\pi}e^{-i\omega t}\epsilon_{\textrm{in}}(0, \omega)\exp\left[-2\pi i\frac{\omega}{c}\int_0^{L}dz \cdot \chi(\omega, z,\Delta_p,\Delta_c)\right],
\end{equation}
where $\epsilon_{\textrm{in}}(z=0, \omega) = \int_{-\infty}^{\infty}dt e^{i\omega t} \epsilon_{\textrm{in}}(z=0, t)$ is the Fourier representation of the initial probe pulse, $L$ is the length of the atomic medium with the optical depth (OD) $d_0 = n_0\lambdabar^2 L$, and $\lambdabar = \lambda/2\pi$.

\begin{figure}[b!]
\def\thefigure{{S4}}
\begin{minipage}[b]{0.35\columnwidth}
\centering
\caption{EIT spectra for the Cs D$_2$ line as a function of the probe detuning for different OD when the control field power is kept constant (1~mW). The intrinsic ground state decoherence is $\gamma_0 = 10^{-3}\Gamma$ and the magnetic field gradient is set to $B_0 = 8$mG.cm$^{-1}$, for a cloud length $L=2.5$~cm. The red solid lines correspond to the present model, while the blue dots are experimental data.}
\vspace{1.8cm}
\label{fig_EIT}
\end{minipage}
\hspace{0.15cm}
\begin{minipage}[b]{0.55\linewidth}
\centering
\includegraphics[width=\textwidth]{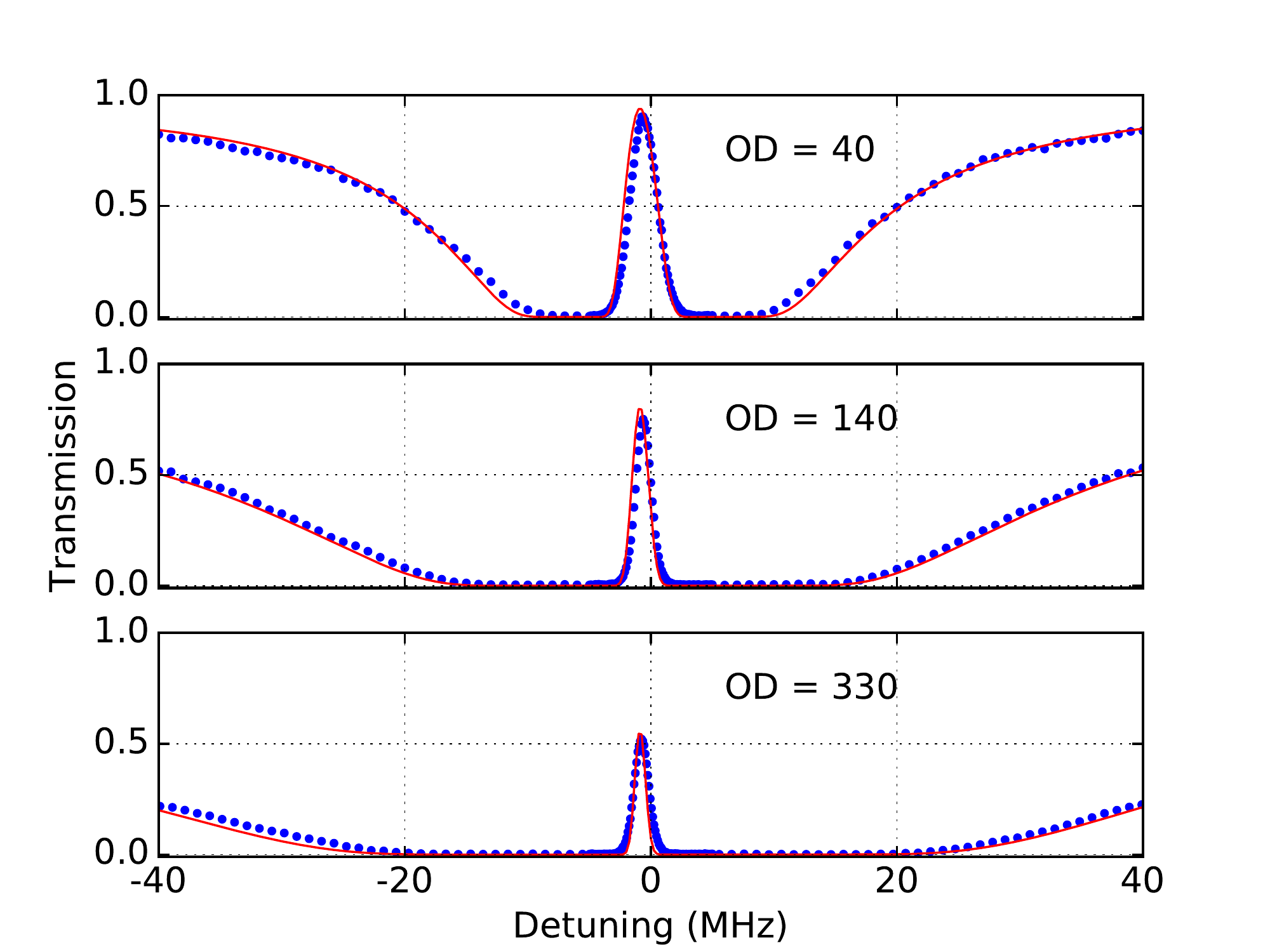}
\end{minipage}
\end{figure}

The transmission spectrum of the probe can thereby be expressed as:
\begin{equation}
T(\omega)  = \exp\left[-4\pi\frac{\omega}{c}\text{Im}\left[\int_0^{L}dz\cdot \chi(\omega, z, \Delta_p,\Delta_c)\right]\right].
\label{transmission}
\end{equation}

The control-induced effective ground state decoherence leads to a reduction of the achievable transparency in the medium. Figure \ref{fig_EIT} gives EIT spectra as a function of the probe detuning, for different ODs but for the same control power, i.e. the same effective decoherence rate. As can be seen, the maximal transmission achieved close to resonance decreases with the OD.

The overall loss during the propagation in the medium gives an upper bound $\eta$ for this storage-and-retrieval efficiency. This bound is given by
\begin{equation}
\eta = \frac{\int_{-\infty}^{\infty}|\epsilon_{\textrm{out}}(z = L, t)|^2dt}{\int_{-\infty}^{\infty}|\epsilon_{\textrm{in}}(z = 0, t)|^2dt}.
\label{memoryefficiency}
\end{equation}

\subsection{Decoherence during the storage duration}
The finite lifetime of the memory leads to a decrease in the overall storage-and-retrieval efficiency. We give here this contribution in the case of residual magnetic fields \cite{Choi2011}. 

Adiabatically switching off the control field coherently converts the probe field into a collective atomic excitation that can be written as:
\begin{equation}
|S(t)\rangle = \frac{1}{\sqrt{N}}\sum_{j = 1}^{N}\sum_{m = -3}^{3}R_m e^{i\phi_{s_mg_m}(t)}|g_{1_m},g_{2_m},...,g_{(j-1)_m},s_{j_m},g_{(j+1)_m},...,g_{N_m}\rangle.
\end{equation} 
The distribution of this collective excitation in the atomic medium depends on the polarizations of the probe and the control fields and it was analysed in details in \cite{Sheremet2010}. The coefficients $R_m = C_{m,1,m+1}^{F_g,1,F_e}/C_{m,1,m+1}^{F_s,1,F_e}$ are ratio of Clebsch-Gordan coefficients for the probe and control fields \cite{Kuzmich,Wang}. The magnetic field gradient for an atom in a state $|g_m\rangle$ and at a position $z$ over time $t$ leads to a phase shift:
\begin{equation}
\Delta\phi_{s_mg_m}(z,t)  = \frac{\upmu_\textrm{B}m(g_{F_s} - g_{F_g})B(z)t}{\hbar}.
 \end{equation}
Assuming a distribution of the atomic density $n_0(z)=n_0 e^{-4z^2/L^2}$, the efficiency $\eta_{s}$ due to the storage time is then evaluated as:
\begin{equation}
\eta_{s} \sim |\langle S(0)|S(t)\rangle|^2 = \left| \frac{1}{N}\sum_{m = -3}^{3}R_m^2\int n(z)e^{i\Delta\phi_{s_mg_m}(z,t)}dz\right|^2 \sim \left| \sum_{m = -3}^{3}R_m^2 e^{-t^2/\tau_m^2} \right|^2,
\label{storage_efficiency}
\end{equation}
where $\tau_m = \frac{2\sqrt{2}}{\upmu_\textrm{B} m(g_{F_s} - g_{F_g})B_0L}$.

\subsection{Numerical simulation: overall storage and retrieval efficiency}

As in the experiment, we consider a probe pulse with a Gaussian temporal profile and a FWHM duration $\tau$ as:
\begin{equation}
\epsilon_{\textrm{in}}(z = 0, t) = \epsilon_0\exp\left[-2\text{ln}2\frac{t^2}{\tau^2}\right].
\end{equation}

In the dynamic storage protocol, at time $T_c$ the control pulse is turned off and the signal pulse is stored inside the atomic medium. Here, for each value of the OD, we choose the power of the control field to obtain a slow-light delay of the pulse $T_d = 2\tau$. In that case, in accordance with \cite{Chen2016}, it can be shown that the writing process has an efficiency close to unity for large OD $(> 50)$, i.e. the leakage is negligible. However at low OD $(< 50)$ the signal pulse cannot be contained entirely and a significant part of the pulse leaks from the medium before $T_c$. A strong leakage is usually observed and limits the efficiency. To take into account the leakage at low OD we change the lower limit of integration of the output pulse in (\ref{memoryefficiency}) to $T_c$. The overall storage-and-retrieval efficiency can also be corrected by the finite efficiency $\eta_{s}$ due to decoherence of the collective excitation during the storage time.

\begin{figure}[t]
\def\thefigure{{S5}}
\begin{minipage}[b]{0.35\columnwidth}
\centering
\caption{Storage-and-retrieval efficiency as a function of the OD using the Cs D$_2$ line. The brown line corresponds to one populated Zeeman sublevel in the ground state $|F_g = 3, m = 3\rangle$, while the blue line corresponds to equally populated Zeeman sublevels. The intrinsic ground state decoherence is $\gamma_0 = 10^{-3}\Gamma$ and the magnetic field gradient amplitude is $B_0 = 8$ mG.cm$^{-1}$, for a cloud length $L=2.5$~cm. The gaussian probe pulse duration is $\tau = 0.5~\upmu$s and the time delay is set to $2\tau$.}
\vspace{1.8cm}
\label{fig_SMOD}
\end{minipage}
\hspace{0.15cm}
\begin{minipage}[b]{0.55\linewidth}
\centering
\includegraphics[width=\textwidth]{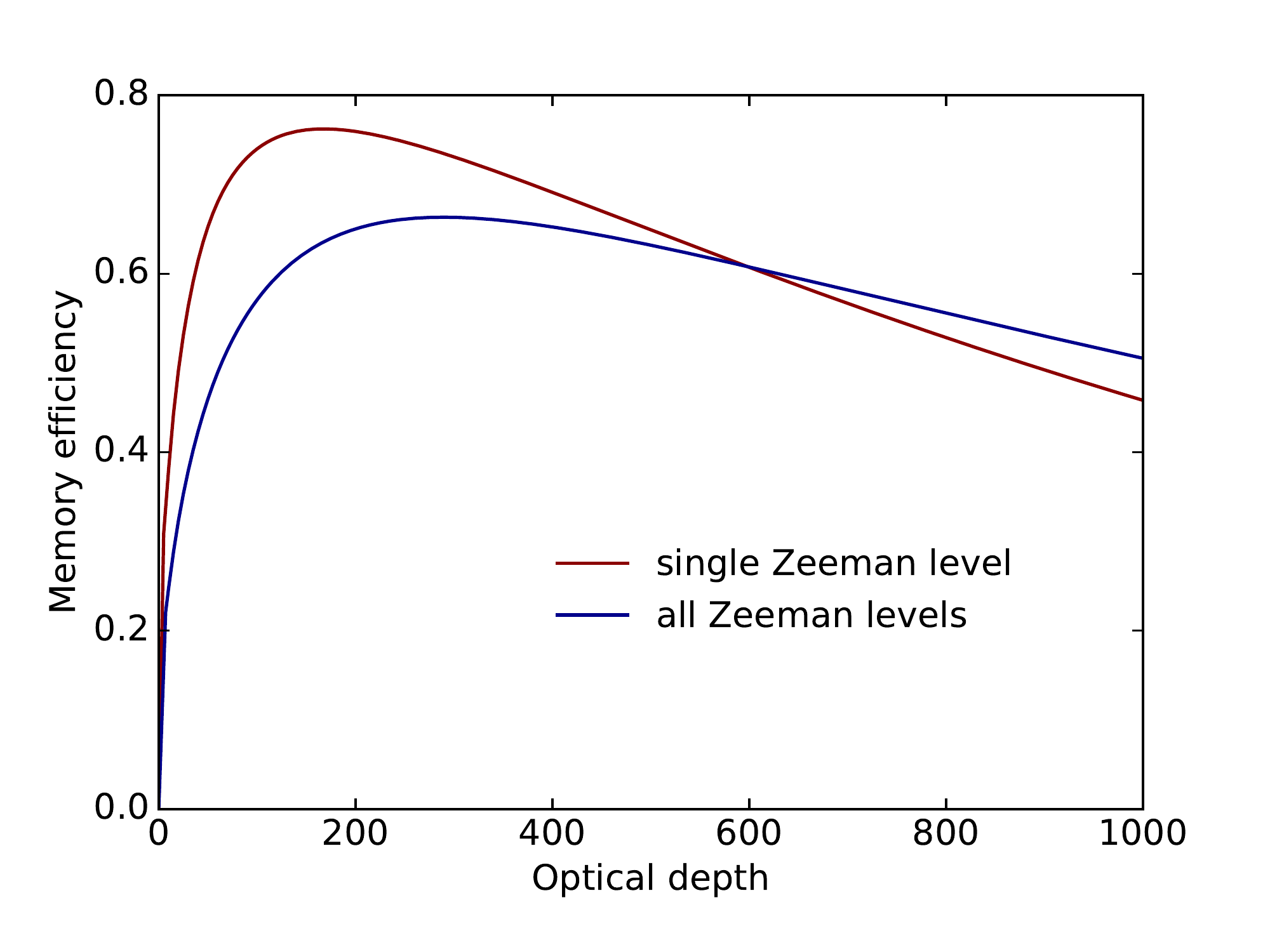}
\end{minipage}
\end{figure}

Figure \ref{fig_SMOD} provides the overall storage-and-retrieval efficiency as a function of OD. The effective ground state decoherence increases with the control field Rabi frequencies. This decoherence becomes more significant in the region of large optical depths due to the large control field power needed. This dependency leads to a reduction of the EIT transparency and, as a result, of the memory efficiency. In our configuration, without Zeeman pumping, the efficiency is limited to about $70\%$. The pumping of atoms to the edge state ($|g\rangle = |6S_{1/2},F = 3, m = 3\rangle$) can lead to a slight increase in the memory efficiency. Our full model enables to compare the two implementations.\\

\end{document}